\documentstyle[11pt,newpasp,psfig,twoside]{article}
\index{instructions}
\index{guidelines}

\def\edcomment#1{\iffalse\marginpar{\raggedright\sl#1\/}\else\relax\fi}
\reversemarginpar

\begin{document}

\newcommand{\psrae}{\mbox{J1847$-$0130}}
\newcommand{\psra}{\mbox{J1847$-$0130 }}
\newcommand{\psrbe}{\mbox{J1718$-$37184}}
\newcommand{\psrb}{\mbox{J1718$-$37184 }}

\title{Two Radio Pulsars with Magnetar Fields}
\author{M. A.~McLaughlin,
 D. R.~Lorimer, A. G. Lyne, \\ M. Kramer, A. J. Faulkner}
\affil{Jodrell Bank Observatory, Macclesfield, UK}
\author{V. M. Kaspi}
\affil{McGill University, Montreal, Canada}
\author{I. H. Stairs}
\affil{University of British Columbia, Vancouver, Canada}
\author{R. N. Manchester, G. Hobbs}
\affil{ATNF -- CSIRO, Epping, Australia}
\author{F. Camilo}
\affil{Columbia University, New York, NY, USA}
\author{A. Possenti \& N. D'Amico}
\affil{Osservatorio Astronomico di Cagliari, Capoterra, Italy}
\begin{abstract}
PSRs~\psra and \psrb have inferred surface dipole magnetic
fields greater than those of any other known pulsars and well above the ``quantum
critical field'' above which some models predict radio emission should not occur.
These fields are similar to those of the anomalous X-ray pulsars (AXPs),
which growing evidence suggests are ``magnetars''. The lack of AXP-like
X-ray emission from these radio pulsars (and the non-detection of radio
emission from the AXPs) creates new challenges for understanding 
pulsar emission physics and the relationship between these classes of
apparently young neutron stars.
\end{abstract}

Both of these pulsars were discovered in the Parkes Multibeam Pulsar Survey
(see e.g. Manchester et al. 2001). PSR~\psra has a spin period of 6.7~s and inferred surface dipole 
magnetic field\footnote{Calculated via the
standard magnetic dipole formula (Manchester \& Taylor 1977) $B=3.2\times10^{19}
\sqrt{P\dot{P}}$ G.}
 of $9.4\times10^{13}$~G. PSR~\psrb
has a period of 3.4~s and magnetic field of
$7.4\times10^{13}$~G.
 The magnetic fields of both 
pulsars are well above the ``quantum critical field'' 
$\simeq 4.4\times10^{13}$~G above which some models predicted radio emission should not occur 
(Baring \& Harding 1998). Both pulsars have average radio luminosities,
indicating that photon splitting does not suppress pair production
at these magnetic field strengths.

In Table~1, we compare the spin-down properties, distances and X-ray luminosities of
the high-field pulsars and the AXPs.
Using archival ASCA data,
we measure an upper limit to the X-ray luminosity of \psra that is lower than the
luminosities of all but one AXP.
 We have analyzed a Chandra observation with \psrb in the field and detect
the pulsar with a soft, thermal spectrum and with an
 X-ray luminosity  much lower than that of any of the AXPs.
H-atmosphere spectral fits
yield a temperature that is consistent with
standard neutron star cooling curves (Kaminker et al. 2001).

\begin{table}
\caption{PSR/AXP Spin Parameters, Distances and Luminosities}
\begin{center}
\begin{tabular}{lrrrr}
\hline  
Name    & $P$ &  $B$ &   $D$   & $L$ (2 -- 10 keV)     \\
& (s) & (10$^{14}$~G) & (kpc) &  ($10^{33}$~ergs  s$^{-1}$) \\
\hline 
1E 1048.1$-$5937 & 6.5 & 5.0 & $\ge$ 2.7 & $\ge$ 5\\
1E 2259+586      &  7.0 & 0.59 &  4$-$7  & $40 - 100$ \\
4U 0142+61       & 8.7 & 1.3 &  $\ge 1.0$ or $\ge 2.7$  & $\ge 10$ or $\ge 70$  \\
RXS J1708$-$40   & 11.0 & 4.6 &  $\sim$ 8  &       $\sim   500$ \\
1E 1841$-$045    & 11.8 & 7.1 &     5.7$-$8.5       &       $20 - 50$ \\
\hline 
PSR~J1718$-$37184 & 3.4 & 0.74 &  $3 - 5$ & 0.0009\\
PSR~J1847$-$0130 & 6.7 & 0.94        &       $6 - 11$       & $< 3 - 8$   \\
\hline 
\end{tabular} 
\end{center}
\end{table}

It is unclear how the high-field pulsars and AXPs can have such similar spin-down
parameters but such different emission properties. One possibility
is that high-field pulsars and AXPs have similar dipole magnetic fields but AXPs have
quadrupole (or higher) components. Finding more high-field radio pulsars is essential for understanding
the relationship between these two populations and constraining the pulsar emission mechanism.
Because of selection effects against the detection of long-period radio pulsars, there may be
many more of these objects than are currently known.
Finally, the discovery of radio emission from these two pulsars shows that
there is no reason, a priori, why the AXPs cannot be radio emitters. While searches for radio emission from the 
currently known AXPs have so far been unsuccessful or unconfirmed (but see Malofeev et al., these
proceedings), this may simply be the result of unfavorable beaming.

\end{document}